\documentclass[useAMS,usenatbib,referee]{mn2e}
\usepackage{graphicx}
   \title{Gemini spectra of 12000K white dwarf stars}

   \author[S. O. Kepler et al.]{S. O. Kepler$^{1}$\thanks{Based on observations obtained at the Gemini Observatory, which is operated by the
Association of Universities for Research in Astronomy, Inc., under a cooperative agreement
with the NSF on behalf of the Gemini partnership: the National Science Foundation (United
States), the Particle Physics and Astronomy Research Council (United Kingdom), the
National Research Council (Canada), CONICYT (Chile), the Australian Research Council
(Australia), CNPq (Brazil) and CONICET (Argentina)},
		B. G. Castanheira$^{1}$,
		A. F. M. Costa$^{1}$,
          and
		D. Koester$^{2}$\\
$^{1}$Instituto de F\'{\i}sica, Universidade Federal do Rio Grande do Sul,
              91501-900  Porto-Alegre, RS, Brazil\\
$^{2}$Institut f\"ur Theoretische Physik und Astrophysik, Universit\"at Kiel,
  24098 Kiel, Germany
             }

\begin{document}
\date{Accepted 2006 Aug 24. Received 2006 July 19}

\pagerange{\pageref{firstpage}--\pageref{lastpage}} \pubyear{2006}
   \maketitle

\label{firstpage}
  \begin{abstract}
We report signal-to-noise ratio ${\mathrm{SNR}} \simeq 100$ optical spectra for 
four DA white dwarf
stars acquired with the
GMOS spectrograph of the 8m Gemini north telescope.
These stars have $18<g<19$ and are around $T_{\mathrm{eff}}\sim 12000$~K,
were the hydrogen lines are close to maximum.
Our purpose is to
test if the effective temperatures and surface gravities
 derived from the relatively low
signal-to-noise ratio ($\langle{\mathrm{SNR}}\rangle\approx 21$) optical spectra acquired by the Sloan Digital
Sky Survey through model atmosphere fitting are trustworthy.
Our spectra range from 3800\AA\ to 6000\AA,
therefore including H$\beta$ to H9.
The H8 line was only marginally present in the SDSS spectra,
but is crucial to determine the gravity.
When we compare the values published by \citet{scot}
and \citet{scot06} with our 
line-profile
(LPT) fits,
the average differences are:
$\Delta T_{\mathrm{eff}}\simeq 320$~K, 
systematically lower in SDSS,
and $\Delta \log g\simeq 0.24$~dex,
systematically larger in SDSS.
The correlation between gravity and effective temperature can
only be broken at wavelengths bluer than 3800\AA.
The 
uncertainties in $T_{\mathrm{eff}}$ are 
60\%  larger, and in
$\log g$ larger by a factor of 4, than the
\citet{scot}
and \citet{scot06} internal uncertainties.
\end{abstract}

\begin{keywords}
   stars -- white dwarfs, techniques: spectroscopic
\end{keywords}

\section{Introduction}

\citet{scot} published the spectra of 2551 white dwarf stars
in the Sloan Digital Sky Survey (SDSS) Data Release 1 (DR1),
increasing the number of spectroscopically identified stars by 
almost 50\%
compared to \citet{McS}. 
\citet{scot06} extended the white dwarf spectroscopic identification
to DR4 with 9316 white dwarf stars reported.
They fit their observed optical
spectra from 3800\AA\ to 7000\AA\ to a grid of synthetic spectra
derived from model atmosphere 
with ML2/$\alpha=0.6$
convective transport in LTE, calculated by Detlev Koester.
Their fits are for the whole spectra and photometry, allowing a 
reflux of 
the models according to a low-order polynomial, to
incorporate effects
of unknown reddening.
The SDSS
spectra have mean signal--to--noise rate SNR(g)$\approx 13$, and
SNR(g)$\approx 21$ for stars brighter than g=19.

   \begin{table}
\begin{minipage}{126cm}
%\centering
      \caption[]{SDSS}
         \label{sdss}
         \begin{tabular}{lccccrccccr}
            \hline
            \noalign{\smallskip}
            Spectra      &  Name & g & $M_g$ & $T_{\mathrm{eff}}$& $\sigma_T$ & $\log g$ & $\sigma_{\log g}$ & Mass & $\sigma_M$ & d \\
      spSpec (MPF)      & (SDSS)&  &  & (K)& (K) & & &($M_\odot$) & ($M_\odot$) & (pc) \\
            \noalign{\smallskip}
            \hline
            \noalign{\smallskip}
51929-0458-188&J030325.22-080834.9&18.74&12.58 &11418.&  119.& 8.500& 0.070&0.925&0.040& 171.\\
52199-0681-079&J233454.17-001436.2&18.34&11.66 &13344.&  321.& 8.140& 0.070&0.699&0.040& 217.\\
51818-0383-111&J232659.23-002348.0&17.50&12.53 &10622.&   47.& 8.330& 0.040&0.815&0.020&  99.\\
51821-0384-008&J233647.01-005114.6&18.29&11.25 &13249.&  247.& 7.860& 0.050&0.544&0.020& 255.\\
            \noalign{\smallskip}
            \hline
         \end{tabular}
\begin{list}{}{}
\item[$^{\mathrm{a}}$] The quoted uncertainties on all tables are the internal uncertainties of the fit
only. 
%The real uncertainty is dominated by the flux and wavelength calibration.
\end{list}
\end{minipage}
   \end{table}
Starting with 
%\citet{koester}, followed by 
\citet{wegner}, 
%\citet{Daou}, \citet{bergeron95}, and \cite{bergeron04}, among others,
most of the white dwarf spectra fits 
for $T_{\mathrm{eff}}$ and $\log g$ are values
derived from the line profiles alone, if a sufficient
set of lines is measured.
\citet{Bergeron95} show SNR$\geq 70$ is required for
uncertainties $\Delta T_{\mathrm{eff}}\leq 300$~K. 
The low order lines are temperature
sensitive, and the higher order lines are pressure -- 
therefore gravity -- sensitive, as they weaken with increasing gravity;
their line profiles are dominated by the quenching of the upper
levels due to the high electronic density. 
%These authors
%recommend fitting only the line profiles of SNR$\geq 70$ spectra, to prevent 
%uncertainties due to flux calibration to be dominant. 
%The H7, H8 and H9 lines, in the violet, are the most sensitive
%to surface gravity because they are produced by electrons at
%higher levels, the most affected by their neighbors. 
However these
lines are also in the region where the atmospheric extinction
is the largest and the CCD detectors the least sensitive.

The SDSS spectra have good flux calibration redwards of 4000\AA, but
have a very low SNR below 4000\AA, even when the spectra extend to H8.
\citet{madej} calculated the mass distribution for the DR1 SDSS
DA sample and found that 
the mean mass increased below $T_{\mathrm{eff}}=12000$~K,
raising a doubt on the \citet{scot} published values.

To test if the SDSS atmospheric values derived from the relatively low SNR spectra
are trustworthy, we obtained ${\mathrm{SNR}}\approx 100$ near 4500\AA\ for 
four DA white dwarf
stars
around $T_{\mathrm{eff}}=12\,000$~K, 
listed in Table~\ref{sdss}.
We calculated the absolute magnitudes 
listed in the table
from the $T_{\mathrm{eff}}$ and $\log g$
obtained in their fits, convolving the synthetic spectra with the g filter 
transmission curve and using the
evolutionary models of \citet{wood} with C/O core, $M_{\mathrm{He}}=10^{-2}\,M_*$, and $M_{\mathrm{H}}=10^{-4}\,M_*$ to estimate their radius.
The distances were then obtained from the distance moduli.

\section{Observations}
We used the Gemini Multi-Object Spectrograph (GMOS) on the 8m Gemini north telescope,
in 1.5" long slit mode, from 3800\AA\ to 6000\AA. We observed with
the B600-G5303 grating
and $2\times 2$ binning, achieving 2.8\AA\ resolution.

The spectra are reduced with the Gemini/GMOS package in IRAF, calibrated with
standard stars observed the same night as the targets, and extinction corrected
using Mauna Kea mean coefficients.
The normal flux and extinction calibration was done with bins of 16\AA, leaving 
undulations in the spectra. As the resolution of our spectra is higher,
we used a 1\AA\ calibration for the primary flux standard star G191-B2B, 
which has been fitted to white dwarf model atmospheres \citep{Bohlin},
and an observation of the star with the
same setup used in our spectra, for a fine calibration.

\section{Models and Fitting}
We employed a synthetic spectra grid similar to that used by \citet{scot},
but extended and denser, to prevent uncertainties in the
fits to dominate the comparison. 
The choice of the ML2/$\alpha=0.6$ parameterization for convection was
demonstrated by \citet{Bergeron95} to give internal consistency with the
temperatures derived in the optical and the ultraviolet, photometry,
parallax and gravitational redshift. It also gave the same mean mass
for his sample with $T_{\mathrm{eff}}$ larger and smaller than 13\,000~K,
while other parameterizations did not. ML2 corresponds to the
\citet{bohm} description of the mixing length theory and
$\alpha=\ell/\lambda_P$ is the ratio of the mixing length to the 
pressure scale height.
%including the $H_2^+$ and $H_2$ quasi-molecular opacities, 
\begin{figure}
   \centering
   \includegraphics[width=\textwidth]{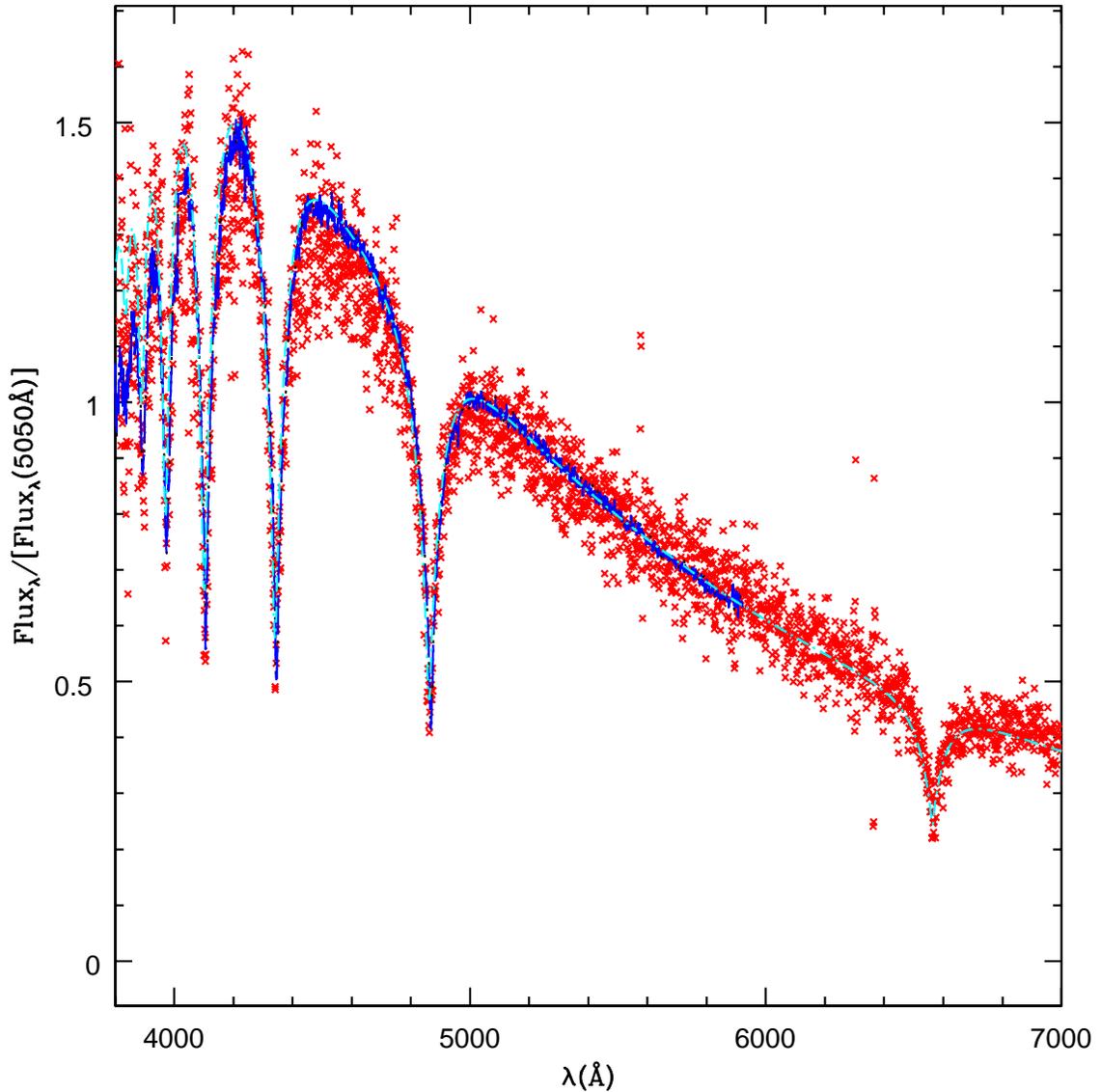}
      \caption{Gemini (solid dark line), SDSS spectra (crosses) of WD J0303-0808, and the model fits (dashed lines).
We plot both the best fit by LPT (Table~2) and the whole spectra fitting (Table~\ref{all}).
              }
         \label{f0303all}
   \end{figure}
\begin{figure}
   \centering
   \includegraphics[width=\textwidth]{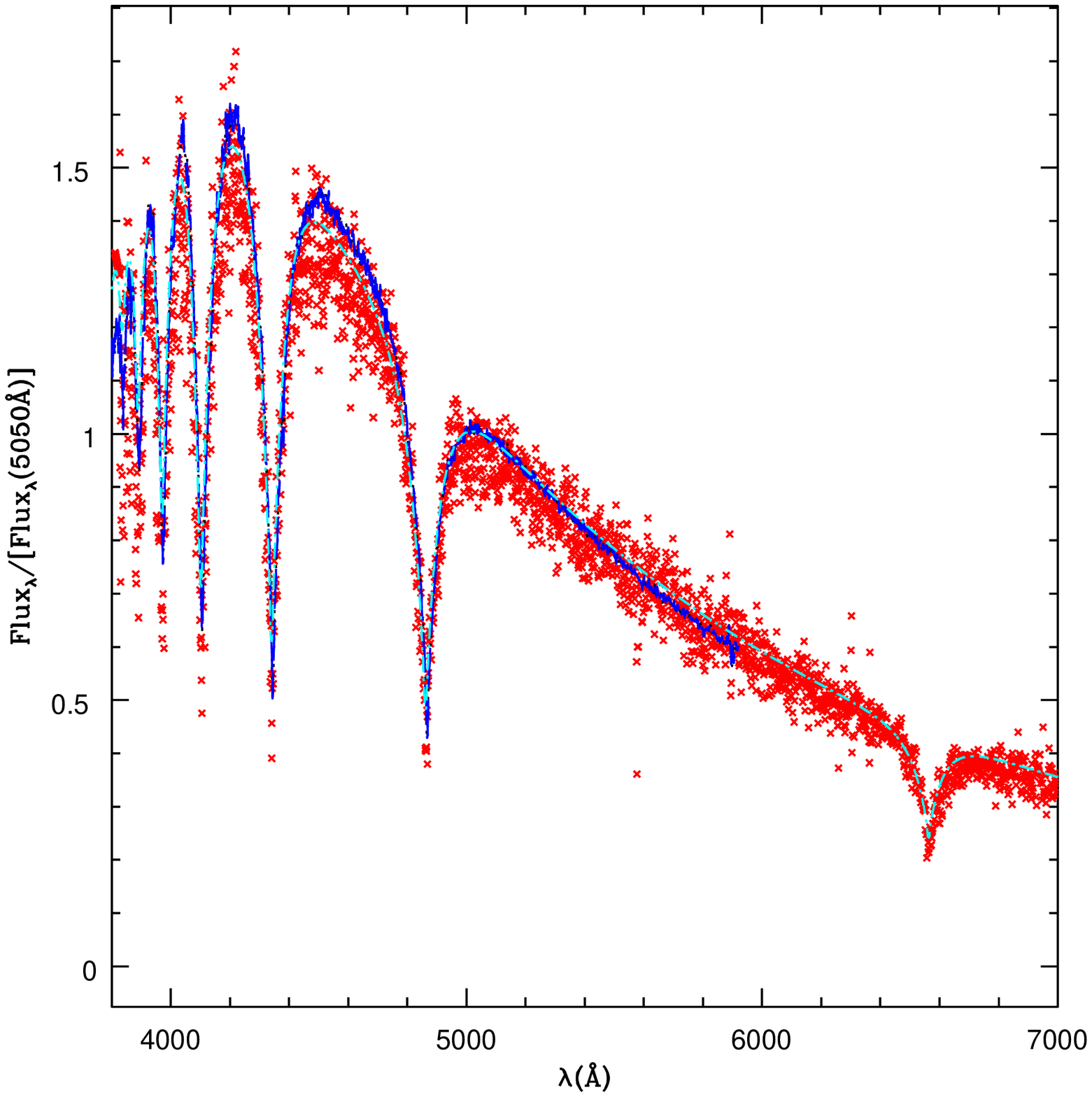}
      \caption{Gemini and SDSS spectra of WD J2334-0014, and the model fits.
              }
         \label{f2334all}
   \end{figure}
\begin{figure}
   \centering
   \includegraphics[width=\textwidth]{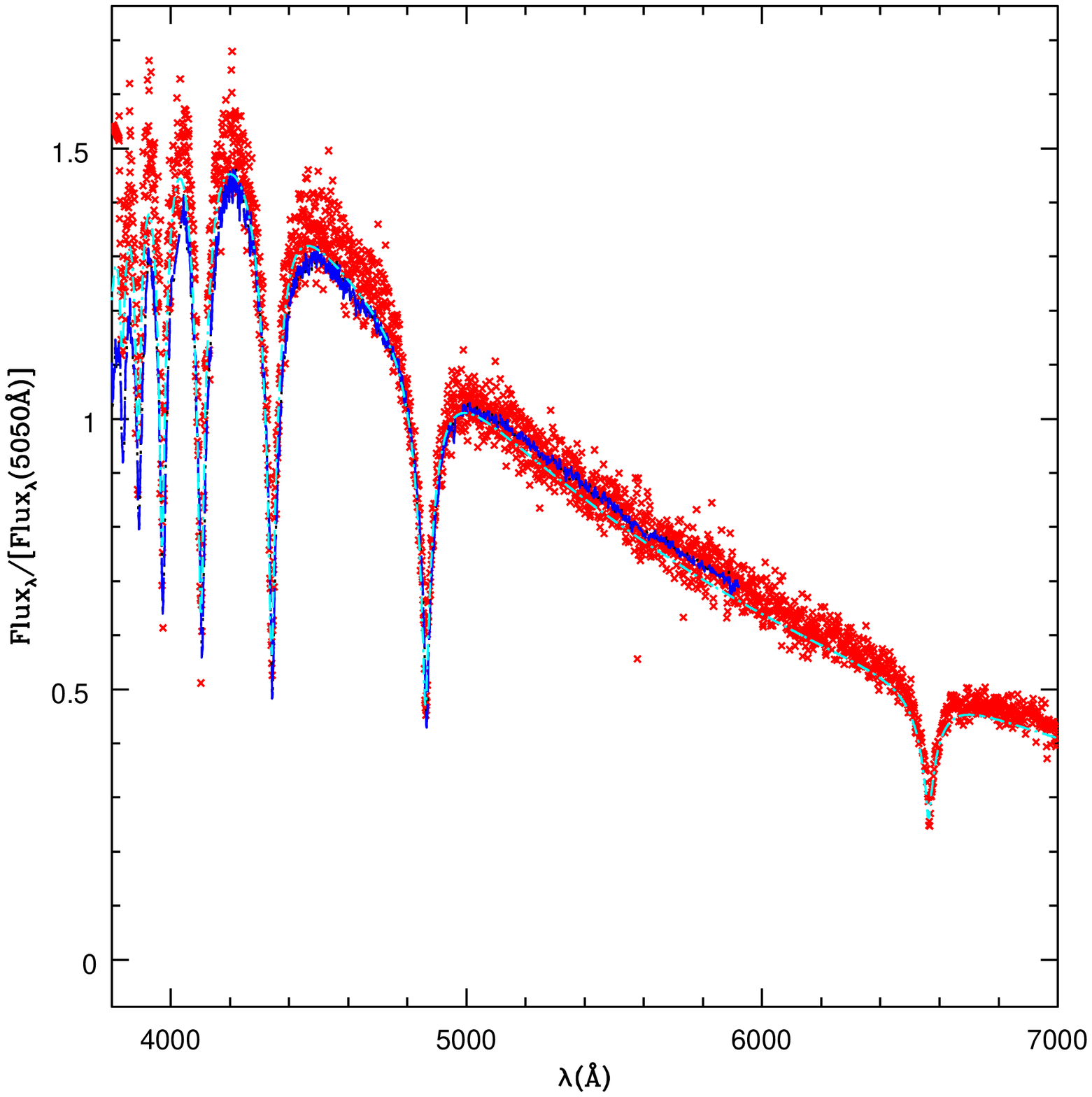}
      \caption{Gemini and SDSS spectra of WD J2326-0023, and the model fits.
              }
         \label{f2326all}
   \end{figure}
\begin{figure}
   \centering
   \includegraphics[width=\textwidth]{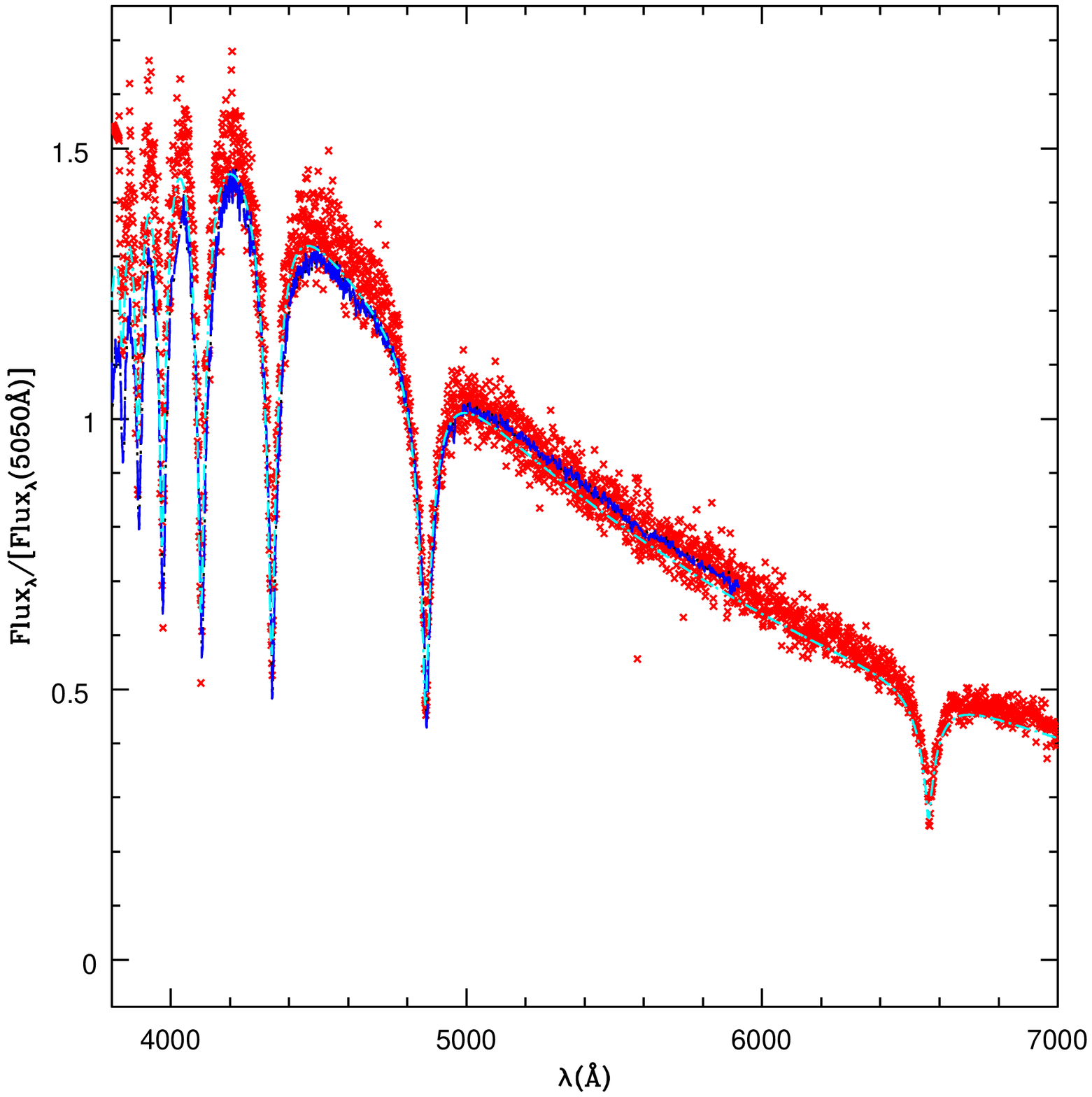}
      \caption{Gemini and SDSS spectra of WD J2336-0051, and the model fits.
              }
         \label{f2336all}
   \end{figure}

\begin{figure}
   \centering
   \includegraphics[width=\textwidth]{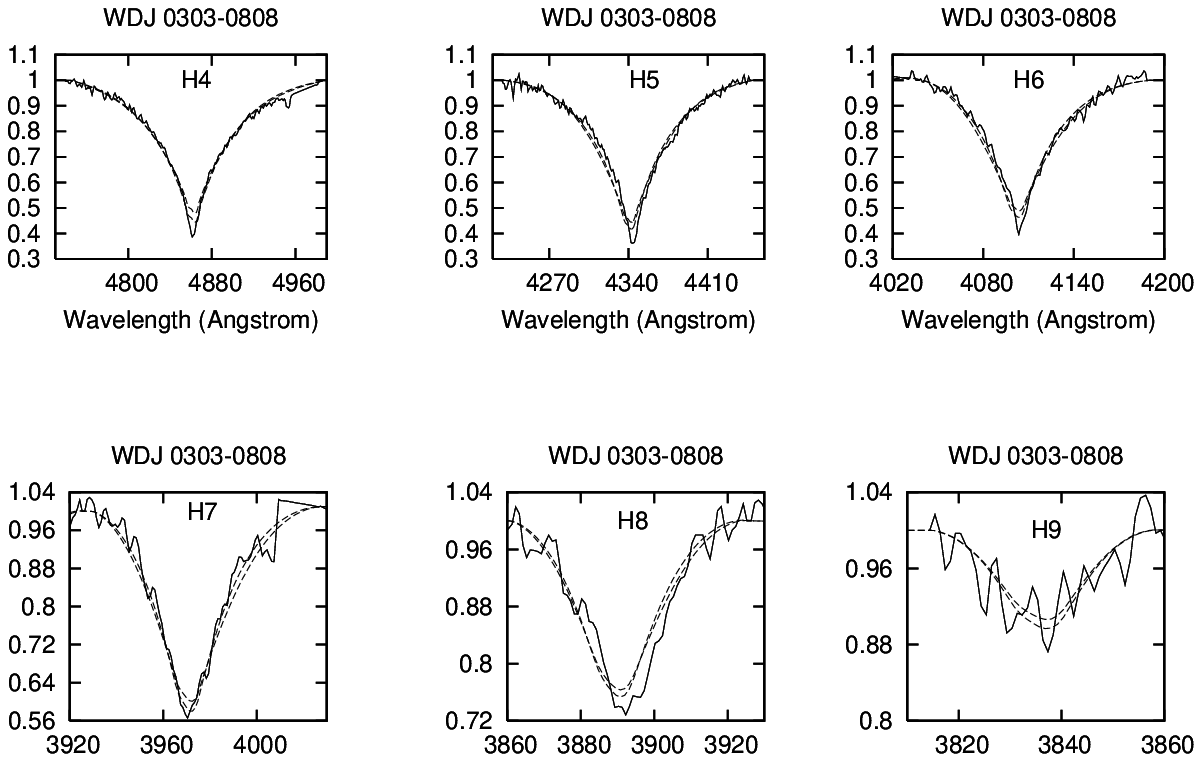}
   \includegraphics[width=\textwidth]{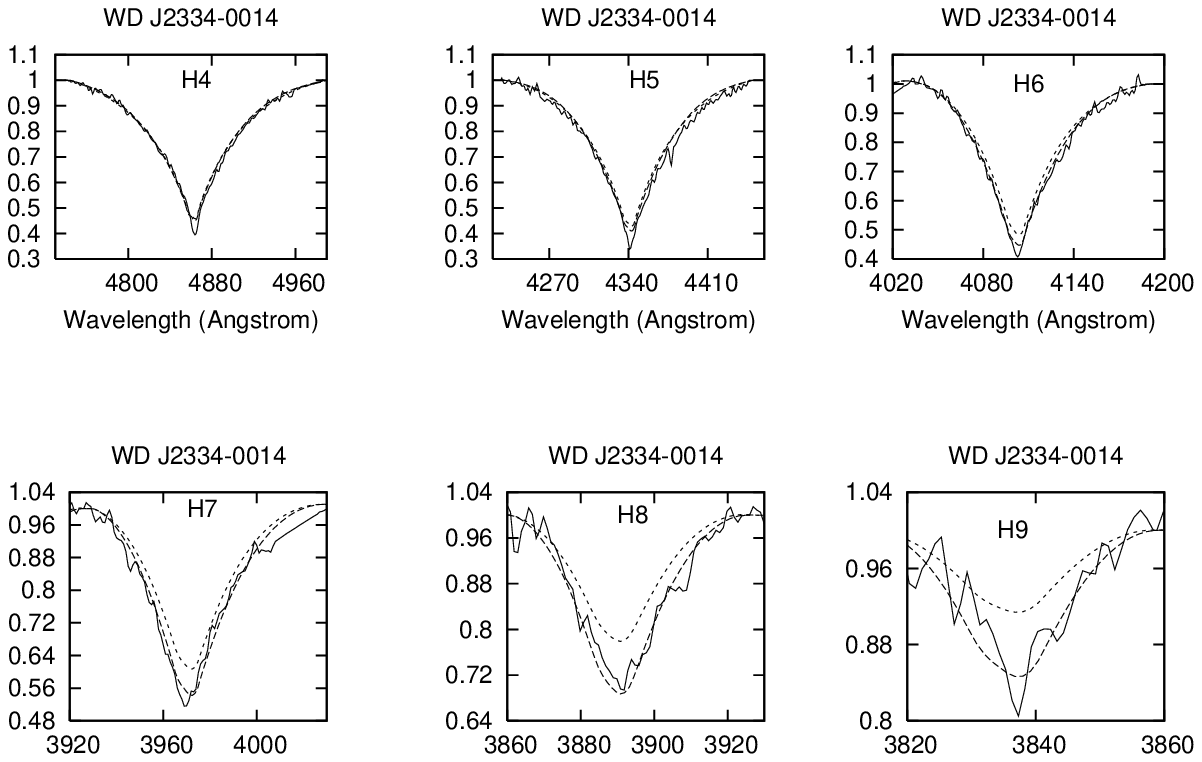}
         \label{flpt}
      \caption{Line profiles in the Gemini spectra of WD J0303-0808, and
WD J2334-0014,
and the models fits. The models fitted by \citet{scot} are the ones
with higher gravities and therefore shallower lines, incompatible with
the observed H9 profiles. Even though the lines were centered before
fitting, we plot here the observed (uncentered) lines.}
   \end{figure}

For line profile technique (LPT) fitting, we normalize both observed spectra and
models to a continuum set at a fixed distance from the line center and
recenter the observed lines to account for radial velocities and
wavelength calibration uncertainties. The synthetic spectra are convolved with
a Gaussian instrumental profile and the whole grid is fitted by least-squares,
weighting all the points equally.
We fitted the spectra both 
whole spectra (Figure~\ref{f0303all}), and line profiles only
(Figure~\ref{flpt}),
to compare with the \citet{scot} values.
The LPT fit does not require spectrophotometric quality data and
is basically immune to flux calibration and reddening uncertainties.
For the whole spectra fitting (all), we normalize the observations and
the models in a region around 5050\AA.
\begin{figure}
   \centering
   \includegraphics[width=\textwidth]{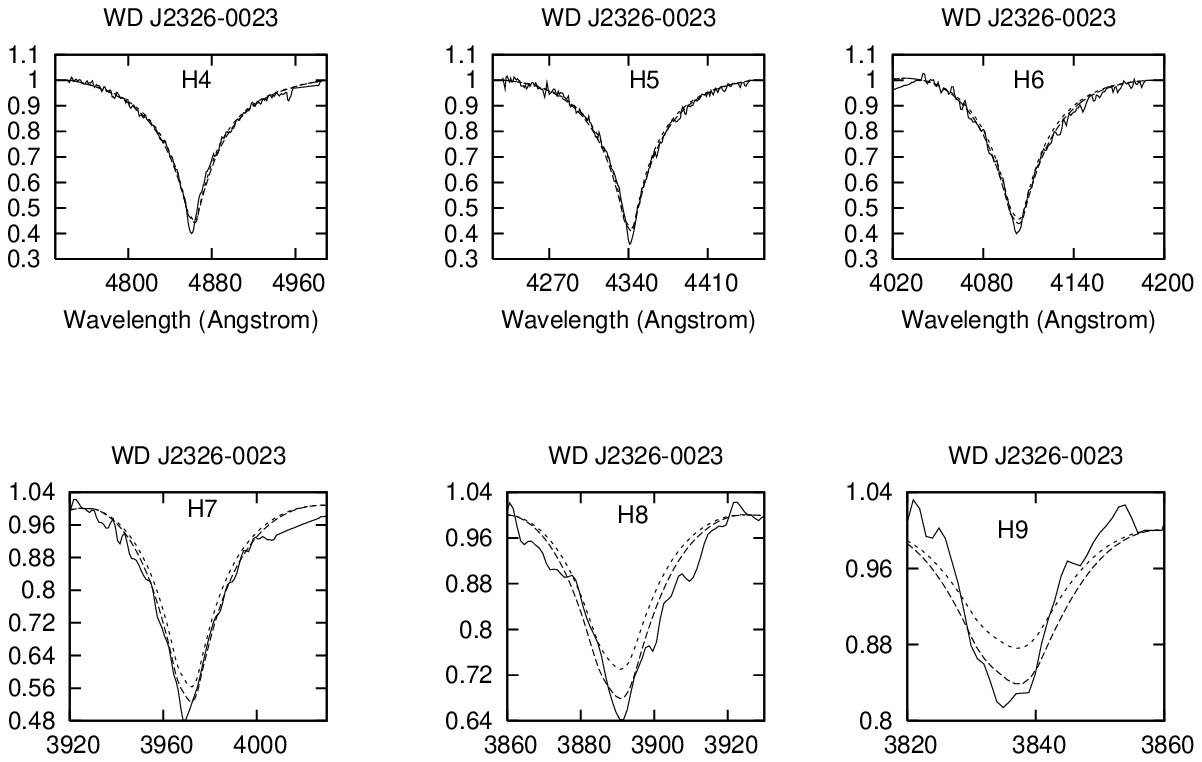}
   \includegraphics[width=\textwidth]{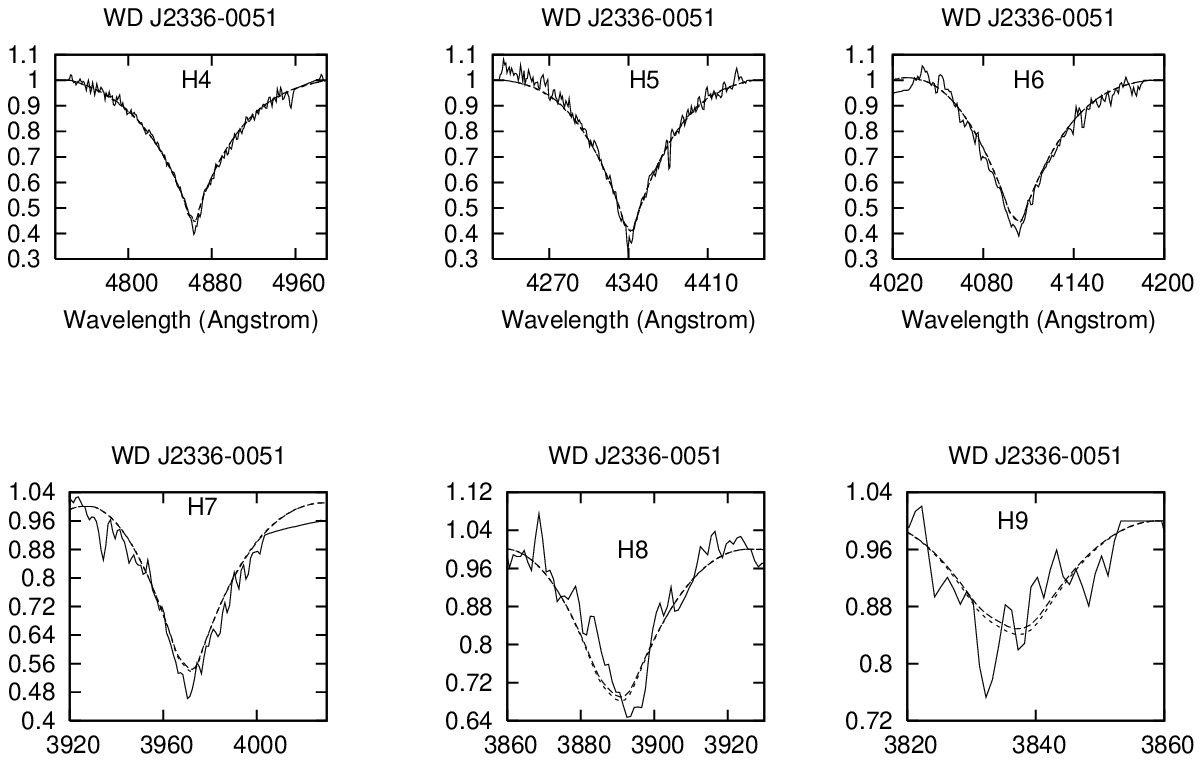}
      \caption{Line profiles in the Gemini spectra of 
WD J2326-0023, WD J2336-0051,
and the models fits. The models fitted by \citet{scot} are the ones
with higher gravities and therefore shallower lines, incompatible with
the observed H9 profiles.
              }
         \label{flpta}
   \end{figure}

%LPT
\begin{table}
\begin{minipage}{126cm}
%\centering
\label{lpt}
 \caption{$T_\mathrm{eff}$ and $\log$ g  with ML2/$\alpha=0.6$ using line profile}
\begin{tabular}{|l|cc|cc|cc|cc|c|}
 {Name} & {T$_\mathrm{eff}$} & {$\sigma_{T_\mathrm{eff}}$}& {$\log g$} & {$\sigma_{\log g}$} & {Mass}&{$\sigma_\mathrm{Mass}$}& {Age(Gyr)}\\
            WD J0303-0808 &   11960. &     160. &  8.305 &  0.017 &  0.791 &  0.003 &  0.724  \\
            WD J2334-0014 &   13543. &     118. &  7.864 &  0.037 &  0.535 &  0.003 &  0.284  \\
            WD J2326-0023 &   10821. &     160. &  8.029 &  0.007 &  0.620 &  0.005 &  0.625  \\
            WD J2336-0051 &   13797. &      21. &  7.712 &  0.009 &  0.461 &  0.005 &  0.223  \\
\end{tabular}
\end{minipage}
\end{table}
%ALL
\begin{table}
\begin{minipage}{126cm}
%\centering
 \caption{$T_\mathrm{eff}$ and $\log$ g  with ML2/$\alpha=0.6$ using the whole spectra}
\label{all}
\begin{tabular}{|l|cc|cc|cc|c|}
  {Name} & {T$_\mathrm{eff}$} & {$\sigma_{T_\mathrm{eff}}$}& {$\log g$} & {$\sigma_{\log g}$} & {Mass}&{$\sigma_\mathrm{Mass}$}&{Age(Gyr)}\\
            WD J0303-0808 &   11423. &     112. &  7.821 &  0.016 &  0.506 &  0.010 &  0.434  \\
            WD J2334-0014 &   13388. &      24. &  7.880 &  0.034 &  0.543 &  0.012 &  0.299  \\
            WD J2326-0023 &   10467. &     404. &  7.800 &  0.012 &  0.492 &  0.012 &  0.535  \\
            WD J2336-0051 &   12192. &     177. &  7.736 &  0.003 &  0.465 &  0.012 &  0.331  \\
\end{tabular}
\end{minipage}
\end{table}
%__________________________________________________________________

We tested \citet{seaton} interstellar reddening proportional to the distances
we measured and found no detectable difference. Even though the distances
are slightly over 100~pc, the SDSS fields were selected perpendicularly
to the galactic disk, so reddening should be low.

To test the fitting method,
we simulated different noise levels added to synthetic spectra.
By Monte
Carlo simulations,
we estimated the average uncertainties in both techniques, LPT and all spectra
fitting. As the signal--to--noise is varying from 100 at 4500\AA\, to less than 
30 at 3800\AA, where the $\log g$ effects are the largest, we report
the simulations up to SNR=60, to 
reinforce the average over wavelength. 
The uncertainties in $T_{\mathrm{eff}}$ are listed in Table~\ref{tabb}, and
we conclude the more trustworthy fits are when we fit all the spectra,
if the uncertainties in flux calibration and interstellar reddening
are not dominant.

\begin{table}
\caption{Uncertainties fitting model with simulated noise spectra.}
\label{tabb}
\begin{tabular}{|c|r|c|r|c|}
SNR & $\sigma_{T_{\mathrm{eff}}}$ (all) &$\sigma_{\log g}(all)$& $\sigma_{T_{\mathrm{eff}}}$ (LPT)&$\sigma_{\log g}$(LPT) \\
10&1550&0.76&1910&0.36\\
20&745&0.16&1105&0.17\\
40&505&0.12&685&0.10\\
60&200&0.05&370&0.07\\
\end{tabular}
\end{table}

\section{Results and Discussions}

   \begin{enumerate}
      \item The effective temperatures derived by \citet{scot} are trustworthy.
The mean difference from SDSS to our high SNR spectra have 
$\Delta T_{\mathrm{eff}} = 320 \pm 200$~K, 
($\Delta T_{\mathrm{eff}} = 370 \pm 230$~K including the variable WD J0303-0808)
systematically lower in SDSS,
where the uncertainties were calculated adding quadratically
the fits internal uncertainties.
      \item The surface gravity uncertainties are underestimated by a factor of 4.
The mean difference from SDSS to our high SNR spectra have 
$\Delta \log g = 0.24 \pm 0.08$,
($\Delta \log g = 0.24 \pm 0.06$ including the variable WD J0303-0808)
systematically larger in SDSS, which corresponds to 
$\Delta {\cal{M}}\simeq 0.13~{\cal{M}}_\odot$ overestimate in mass.
   \end{enumerate}

The main difficulty in these fits is the correlation between the
derived $T_{\mathrm{eff}}$ and $\log g$ --- a small increase in 
$T_{\mathrm{eff}}$ can be compensated by a small decrease in
$\log g$ --- as for WD J0303-0808, for which the model with
$T_{\mathrm{eff}}=12000$~K, $\log g=8.0$, from our LPT fit, differs from
the whole spectra best fit $T_{\mathrm{eff}}=11400$~K, $\log g=8.3$ only below
3800\AA, where we have no measured flux.

The four stars reported here were classified as Not-Observed-to-Vary (NOV) by
\citet{Mukadam04}, but \cite{barbara06} report WD J0303-0808 is, in fact, a
low amplitude pulsator. Pulsation does introduce a real variation of
the measured effective temperature of 50 to 500~K, depending on the
real amplitude of the pulsation, during a cycle \citep{Kepler84}.

A systematic increase in the measured gravity for white dwarfs at effective temperatures
lower than 12\,000~K has been observed for more than a decade, and for the SDSS spectra
has been reported by \citet{scot}, \citet{madej}, and \citet{scot06}. Our results indicate
that such an increase, in the region of Balmer line maximum,
$14000~\mathrm{K} \geq T_{\mathrm{eff}} \geq 11000$~K, is due to the projection of the
$T_{\mathrm{eff}}$--$\log g$ real correlated solution onto a smaller
$T_{\mathrm{eff}}$ range. Such projection would also explain why 
\citet{Mukadam04} and \citet{Mullally} find a narrower ZZ Ceti instability
strip than \citet{Bergeron04} and \citet{Gianninas}, but contaminated
by non-variables. 

Acknowledgments: Gemini GN-2005B-Q-67.

\bsp

\label{lastpage}

\end{document}